\documentclass[12pt]{article}
\usepackage{jheppub}
\usepackage{amssymb,amsfonts}

\usepackage{wrapfig}
\usepackage{epsfig}
\usepackage[T1]{fontenc}
\usepackage[utf8]{inputenc}
\usepackage{lmodern}
\usepackage{float}
\usepackage{placeins}
\usepackage{dsfont}
\usepackage{arydshln}

\usepackage[dvipsnames]{xcolor}
\usepackage{tikz}

\def\be{\begin{eqnarray}}
\def\ee{\end{eqnarray}}
\newcommand{\nn}{\nonumber}
\newcommand\para{\paragraph{}}
\newcommand{\ft}[2]{{\textstyle\frac{#1}{#2}}}
\newcommand{\eqn}[1]{(\ref{#1})}

\newcommand{\ra}{\rightarrow}
\newcommand{\ket}[1]{|#1\rangle}
\newcommand{\bra}[1]{\langle#1|}

\def\Dslash{\,\,{\raise.15ex\hbox{/}\mkern-12mu D}}
\def\Dbarslash{\,\,{\raise.15ex\hbox{/}\mkern-12mu {\bar D}}}
\def\delslash{\,\,{\raise.15ex\hbox{/}\mkern-9mu \partial}}
\def\delbarslash{\,\,{\raise.15ex\hbox{/}\mkern-9mu {\bar\partial}}}
\def\pslash{\,\,{\raise.15ex\hbox{/}\mkern-9mu p}}
\def\calDslash{\,\,{\raise.15ex\hbox{/}\mkern-12mu {\cal D}}}

\newcommand{\Z}{{\mathds Z}}

\newcommand{\e}{\,{\rm e}}

\def\lae{\mathrel{\mathop{\smash{\lower .5 ex \hbox{$\stackrel<\sim$}}}}}
\def\lae{\mathrel{\mathop{\smash{\lower .5 ex \hbox{$\stackrel>\sim$}}}}}

\title{Dynamics of the Fermion-Rotor System}

\author[a]{Vazha Loladze}
\author[b,c]{Takemichi Okui}
\author[d]{and David Tong}

\affiliation[a]{Rudolf Peierls Centre for Theoretical Physics, University of Oxford, Parks Road, Oxford, OX1 3PU, UK}
\affiliation[b]{Department of Physics, Florida State University, Tallahassee, FL 32206, US}
\affiliation[c]{Higher Energy Accelerator Research Organization (KEK), Tsukuba 305-0801, Japan}
\affiliation[d]{Department of Applied Mathematics and Theoretical Physics \\ University Cambridge, CB3 0WA, UK}

\emailAdd{vazha.loladze@physics.ox.ac.uk}
\emailAdd{tokui@fsu.edu}
\emailAdd{d.tong@damtp.cam.ac.uk}

\abstract{We explore the dynamics of the  fermion-rotor system, a simple impurity model in $d=1+1$ dimensions consisting of a collection of purely right-moving fermions interacting with a quantum mechanical rotor localised at the origin. This was first introduced by Polchinski as a toy model for monopole-fermion scattering and is surprisingly subtle, with ingoing and outgoing fermions carrying different quantum numbers. We show that the rotor acts as a twist operator in the low-energy theory, changing the quantum numbers of excitations that have previously passed through the origin to ensure scattering consistent with all symmetries.  

We further show how generalisations of this model with multiple rotors and unequal charges can be viewed as a UV-completion of boundary states for chiral theories, including the well-studied 3450 model. We compute correlation functions between ingoing and outgoing fermions, and show that fermions dressed with the rotor degree of freedom act as local operators and create single-particle states, generalizing an earlier result obtained in a theory with a single rotor and equal charges. Finally, we point out a mod 2 anomaly in these models that descends from the Witten anomaly in 4d.}

\begin{document}
\pagestyle{plain} \setcounter{page}{1}
\newcounter{bean}
\baselineskip16pt \setcounter{section}{0}

\maketitle


\section{Introduction}

The fermion-rotor system is a deceptively simple model in $d=1+1$ dimensions constructed long ago by Polchinski \cite{joe}. The model consists of a collection of purely right-moving  massless fermions, interacting with an impurity which takes the form of a  quantum mechanical rotor localised at the origin of space.  The subtlety in the model arises because of a quantum anomaly which ensures that  ingoing fermions carry different quantum numbers from the outgoing fermions. This makes it challenging to see how consistent scattering can be achieved while preserving all symmetries.

 \para
 This challenge was recently overcome in 
\cite{lt}. There it was shown how the outgoing fermions can carry quantum numbers that differ from those in the naive Fock space of the theory. This is possible  because the fermions are dressed with an operator associated to the rotor, evaluated at the time in the past when the fermion passed through the origin. At first glance, dressing a fermion far from the origin with the rotor would seem to give rise to something non-local. Yet this is not what happens, at least not in 2-point correlation functions. The net result is that, surprisingly, fermions propagate through the rotor as free single-particle states.
 
 \para
 Polchinski's motivation for designing  the fermion-rotor system was to have a simple toy model that captures the physics of massless fermions scattering off magnetic monopoles in $d=3+1$ dimensions.  It is well known that fermion-monopole scattering can violate symmetries which suffer an anomaly,  a phenomenon known as the Callan-Rubakov effect \cite{c,r}.  But Callan also pointed out that there are situations where a seeming paradox arises, with no possible outgoing state consistent with the quantum numbers of the ingoing states \cite{callan}.   Importantly, this paradox  exists for monopoles in the Standard Model and there is no consensus in the community as to what happens when, say, a right-handed electron scatters off a magnetic monopole.   In recent years, there has been renewed interest in this paradox, with a number of different resolutions proposed \cite{lt, pancake,csaki,csaki2,brennan,brennanplus,brennan2,fock,me,me2,hook,valya}.

\para
The fermion-rotor system  captures the physics of fermions scattering off 't Hooft-Polyakov monopoles in $d=3+1$ dimensions. This arises as follows: first, the monopole scattering paradox exists only for s-wave states (or, more generally, for lowest angular momentum states) which means that the question can be  framed purely in  $d=1+1$ dimensions, rather than the full $d=3+1$ dimensions. Next, the outgoing modes are ``unfolded'', so instead of thinking of ingoing and outgoing modes as right- and left-movers, living on a half-line with boundary, we instead consider purely right-moving fermions  that pass through an  impurity. The impurity rotor degree of freedom is playing the role of the dyonic collective coordinate of the 't Hooft-Polyakov monopole, whose excitations endow the monopole with electric charge \cite{jz}. In this way,  at low energies and weak coupling, the scattering of fermions off monopoles reduces to the fermion-rotor problem. 

\para
Many of the other proposed resolutions to the monopole scattering paradox instead focus on singular Dirac monopoles, or 't Hooft lines in their modern incarnation. 
Here the paradox has a slightly different flavor: there is no longer a dynamical rotor degree of freedom, and the problem becomes one of what boundary conditions should be imposed on the singular 't Hooft line, and how to interpret the resulting physics.  It is natural to ask whether the solution to the fermion-rotor problem presented in \cite{lt} makes contact with any of the other approaches to the monopole-fermion paradox. The purpose of this paper is to answer this in the affirmative. We will show that the fermion-rotor system can be viewed as a UV completion of the boundary conformal field theory (BCFT) approach discussed in \cite{me,me2}. 

\para
One  of the general lessons of BCFT is that the kinds of local boundary conditions that can be placed on a quantum field theory are much more varied than the semi-classical Dirichlet or Neumann conditions that are familiar. Affleck and Sagi were the first to apply the powerful technology of BCFT to the problem of monopole scattering \cite{sagi}. This was explored further in  \cite{me,me2}, where it was argued that the boundary conditions imposed by the monopole act as a portal into a twisted sector of the Hilbert space.  This generalises an earlier result of  Maldacena and Ludwig, who looked at a model with eight Majorana fermions that exhibits $SO(8)$ triality \cite{mlud}.  Similar ideas were also put forward in \cite{fock}.

\para
The idea that scattering takes place into a twisted Hilbert space resonates with the solution of the fermion-rotor problem given in \cite{lt}, where it was shown that there exists a different sector of outgoing fermions, carrying unexpected quantum numbers. Our first goal in this paper is to make this analogy precise. In Section \ref{frsec}, we introduce the fermion-rotor system and show that, at low energies, the rotor degree of freedom acts as a twist operator, giving rise to the same physics seen in \cite{me,me2}.

\para
In Section \ref{chargesec}, we further explore the relationship between the fermion-rotor system and BCFT results. We generalise the fermion-rotor system to include multiple rotors, coupled with different charges. This generalisation provides a rotor manifestation of a large class of (unfolded) chiral theories, including the 3450 model, and allows us to make contact with a class of boundary states whose properties were studied in  \cite{philip1} (see also \cite{costas,ryu,philip2}).

\para
 In section \ref{corrsec}, we compute correlation functions in this generalised system, following the method used in \cite{joe,lt}. As in the simplest system in \cite{joe, lt}, a naive regularisation of IR divergence misses an important physics associated with vacuum degeneracy, but it can be rectified using cluster decomposition techniques.

\para
Finally, in Section \ref{mzsec}, we describe an anomaly in this system. The fermion-rotor system comprises of chiral fermions, interacting with a dynamical impurity that looks, in some ways, like a localised chemical potential. One might wonder if it suffers an anomaly. We argue that it does, albeit only a mod 2 anomaly. In the fermion-rotor description, this manifests itself as a Grassmann-odd vev. Another way to exhibit the mod 2 anomaly  is to compactify the system on a spatial circle where, for certain couplings to the rotor, the system has an odd number of Majorana zero modes.

\begin{figure}[htb]
\begin{center}
\epsfxsize=5.2in\leavevmode\epsfbox{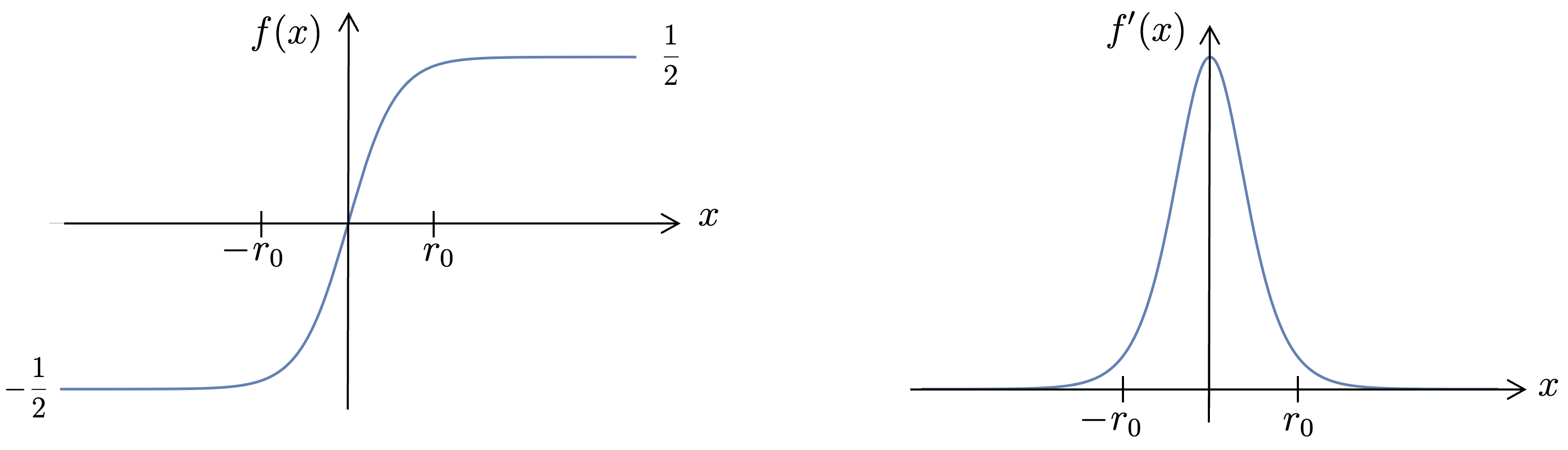}
\end{center}
\caption{The function $f(x)$ is a regularised step function, while $f'(x)$ is a regularised delta-function, with the length scale $r_0$ acting as the regulator.}\label{fxfig}
\end{figure}
\noindent

\section{The Fermion-Rotor System}\label{frsec}

The original Polchinski rotor  model has $N$ right-moving Weyl fermions, $\psi_i(t,x)$ with $i=1,\ldots,N$, coupled to a rotor degree of freedom $\alpha(t)$, with action
\be S = \int d^2x \ i\sum_{i=1}^N\psi_i^\dagger\big(\partial_++ i\alpha(t) f'(x)\big) \psi_i + \int dt \ \frac{I}{2} \dot{\alpha}^2\ .\label{polact}\ee
Here the chiral derivative  $\partial_+=\partial_t+\partial_x$ reflects the fact that our fermions are purely right-moving,  $I$ is the moment of inertia for the rotor, while $f'(x)$ is a fixed background that should be viewed as a regularised delta-function sitting at the origin, as shown in Figure \ref{fxfig}. We take this to be an even function, normalised as
\be \int_{-\infty}^{\infty} dx\ f'(x) = 1\ .\ee
In Section \ref{chargesec}, we will generalise the rotor model to include different charges, and multiple rotors. But for now we summarise some of the key features of this simple theory.

\para
The classical action has a global $SU(N)$ symmetry, rotating the fermions $\psi_i$. This $SU(N)$ is non-anomalous and survives in the quantum theory.
Classically, the action also enjoys a global $U(1)$ symmetry, acting a
\be 
\psi_i(t,x) \ra e^{i\beta} \psi_i(t,x)
\,.\label{notsym}
\ee
However, this is not a symmetry of the quantum theory as it suffers an anomaly. To see this, note that the rotor coupling could be viewed as a gauge field $A_1(t,x) = \alpha(t) f'(x)$, albeit one that is slightly unusual because $\alpha(t)$ is dynamical while $f'(x)$ is fixed. This corresponds to a field strength $F_{01} =\dot{\alpha} f'(x)$. The would-be current associated to this symmetry is $J=\sum_i \psi^\dagger_i\psi_i$, but the anomaly means that this is not-conserved, instead obeying
\be 
\partial_+ J(t,x)  = \frac{N}{2\pi} \,\dot{\alpha}(t) f'(x)
\,.\label{anomaly}
\ee
The rotor model does, however, have a different global $U(1)$ symmetry that is non-anomalous and survives in the quantum theory. This acts as
\be 
\psi_i(t,x) \ra e^{i\beta f(x)} \psi_i(t,x) 
\ \ \ {\rm and}\ \ \ 
\alpha(t) \ra \alpha(t) -\beta
\,.\label{symf}
\ee
The presence of the function $f(x)$ in the exponent is what makes things interesting. This is a non-anomalous symmetry providing that we take $f(x)$ to be odd, with asymptotic behavior $f(x) \ra \pm \ft12$ as $x\ra \pm\infty$. It is this requirement that fixes the integration constant in $f(x)$ when we derive it from integrating $f'(x)$. (The action is in terms of $f'(x)$, not $f(x)$.) But the presence of $f(x)$ in the $U(1)$ transformation means that the ingoing fermions to the left of the rotor carry (up to a factor of $1/2$) charge $-1$ under this $U(1)$, while the outgoing fermions to the right of the rotor carry charge $+1$. And therein lies the rub.

\para
The upshot is that incoming ($x<0$) fermions $\psi_i(x)$ sit in the representation ${\bf N}_{-1}$ under the $SU(N)\times U(1)$ symmetry, while outgoing ($x>0$) fermions $\psi_i(x)$ sit in the representation ${\bf N}_{+1}$. How, then, can consistent scattering be achieved conserving both symmetries?

\para
There is an obvious answer to this question that is wrong. This follows from the observation that the $U(1)$ symmetry \eqn{symf} shifts $\alpha$, and 
so the corresponding Noether charge also depends on the rotor in addition to the fermions, 
and it is given by
\be
Q = I\dot\alpha(t) + \int_{-\infty}^\infty dx\, f(x) \, J(t,x)   
\,.\label{U(1)charge}
\ee
This means that a spinning rotor carries charge $Q_{\rm dyon} = I\dot\alpha$ which, following the connection to monopoles, we refer to as the dyon charge. This suggests that as a fermion passes through the origin, it deposits charge $-2$ on the rotor. However, this cannot happen at low energies. The energy of the excited 
rotor is $E_{\rm dyon} = I\dot\alpha^2/2 = Q_{\rm dyon}^2/2I$. A charge $Q_{\rm dyon} \sim {\cal O}(1)$ entails an energy $E_{\rm dyon} \sim 1/I$. If we send in the massless fermion with an energy $E\ll 1/I$ then the resulting excited rotor 
must be  highly off-shell and cannot last longer than a time scale $\sim I \ll 1/E$. Indeed,  as we will review below, any charge on the rotor quickly decays. So what does happen?

\para
The correct answer is much less obvious and  was given in \cite{lt}: the fermion with quantum numbers ${\bf N}_{-1}$ propagates freely through the rotor impurity, and continues as an outgoing state. This is surprising because, as we've seen, naively there is no propagating state with quantum numbers ${\bf N}_{-1}$ to the right of the rotor. Instead,  the operator $\psi_i^\dagger(t,x)$ creates a state with quantum number ${\bf N}_{+1}$.  The key result of \cite{lt} is that the operator 
\be {\cal O}_i(t,x) = e^{i\alpha(t-x)}\,\psi_i(t,x) \ee
does the job, carrying the correct quantum numbers ${\bf N}_{-1}$. At first glance, ${\cal O}_i(t,x)$ appears to be a non-local operator, with the fermion at the point $(t,x)$ dressed with the rotor degree of freedom evaluated at the retarded time $t-x$ when the fermion interacted with the impurity. Nonetheless, as shown in \cite{lt}, there is no hint of this non-locality when the operator is evaluated in correlation functions. Indeed, the simplest such correlation function is 
\be 
\bigl\langle e^{i\alpha(t-x)}\psi_i(t,x) \, \psi_j^\dagger(t',x') \bigr\rangle  
= \delta_{ij} G_0(t-t',x-x')
\ee
where $x'<0$ and $x>0$, so this corresponds to an ${\bf N}_{-1}$ fermion sent in from the left, passing through the rotor. The right-hand side $G_0(t-t',x-x')$ is simply the propagator for a free chiral fermion. We review the derivation of this result in Section \ref{corrsec}.

\subsection{The Rotor and the Twist}

We now look more closely at the rotor operator $e^{i\alpha(t)}$. We will show that this obeys the equal-time exchange relations appropriate for a twist operator. 

\para
To do this, we first solve the quantum equations of motion for the fermion-rotor system. The equation of motion for the rotor is
\be I\ddot{\alpha}(t) =- \int_{-\infty}^{+\infty} dx\, f'(x) \, J(t,x) 
\,.\label{rotoreom}
\ee
The current $J$ is, in turn,  related to the rotor degree of freedom through the anomaly equation \eqn{anomaly}. 
%
%
We can integrate \eqn{anomaly} once to give
\be J(t,x) = J(t-x+x_0,x_0) + \frac{N}{2\pi} \int_{x_0}^x dx'\, f'(x') \, \dot{\alpha}(t-x+x')\label{withx0}\ee
where, at this stage, $x_0$ is an arbitrary integration constant. 

\para
Setting the integration constant to the two values $x_0=\pm r_0$, with $r_0$ the delta-function cut-off,  provides different insights into the nature of the rotor dynamics. We start by setting $x_0=-r_0$, a point just to the left of the rotor.  We then take the limit $r_0\ra 0$, so that $f'(x') \ra \delta(x')$. This gives
\be J(t,x) = \tilde{J}_{\rm in}(t-x) + \frac{N}{2\pi} \theta(x) \dot{\alpha}(t-x)\ .
\label{eq:currentin}
\ee
Here  $\theta(x)$ is the usual Heaviside step function, and we've introduced the incoming current immediately to the left of the rotor,
\be \tilde{J}_{\rm in}(t) = \displaystyle{\lim_{r_0\ra 0}}\, J(t-r_0,-r_0) \ .\ee
If we now substitute the  expression \eqn{eq:currentin} into the equation of motion \eqn{rotoreom}, we get
\be I\ddot{\alpha}(t)  = -\int_{-\infty}^\infty dx\, f'(x) \,\tilde{J}_{\rm in}(t-x) - \frac{N}{2\pi} \int_0^\infty dx\, f'(x)\,\dot{\alpha}(t-x)  \ee
Again, we take the $f'(x) \ra \delta(x)$ limit in both terms here but, in the second term, we pick up a factor of $1/2$ because the integral is only along the half-line. The end result is that the dynamics of the rotor is governed by
\be I\ddot{\alpha}(t) 
=-\tilde{J}_{\rm in}(t) -\frac{N}{4\pi} \dot{\alpha}(t)
\,.\label{friction}
\ee
The result of the anomaly is manifest here: it appears in the second term above which acts as  effective friction term for the rotor. Finding a friction term arising in a Hamiltonian system is rather unusual.  But, as we've seen, the friction term doesn't come from the classical equations of motion. Instead, it comes from the anomaly. Moreover, it's clear what the resulting physics is: any excitation of the rotor caused by the incoming current $\tilde{J}_{\rm in}(t)$ is quickly damped in a time scale $t\sim 4\pi I/N$. The energy, and any quantum charge, stored in the rotor must be emitted in outgoing fermions.

\para
We can get a different perspective by returning to \eqn{withx0} and setting the integration constant to be $x_0=r_0$, the right-most cut-off of the delta-function. This will relate the rotor degree of freedom to the outgoing current, just to the right of the rotor,
\be \tilde{J}_{\rm out}(t) = \displaystyle{\lim_{r_0\ra 0}}\, J(t+r_0,r_0) \ .
\ee
Repeating the steps above, including taking the limit $r_0\ra 0$, means that the expression for the current \eqn{eq:currentin} is replaced by
\be J(t,x) = \tilde{J}_{\rm out}(t-x) - \frac{N}{2\pi} \theta(-x) \dot{\alpha}(t-x)\ .
\label{eq:current}
\ee
Substituting this into the equation of motion \eqn{rotoreom}, and taking the $r_0\ra 0$ limit,  now gives the result
\be 
I\ddot{\alpha}(t) 
= - \tilde{J}_{\rm out}(t) + \frac{N}{4\pi} \dot{\alpha}(t)
\,.\label{eom-in-terms-of-Jtilde}
\ee
Naively the $+$ve sign in the second term makes it look as if the rotor is speeding up! But that's misleading because an excited rotor will induce an outgoing current $\tilde{J}_{\rm out}(t)$, so the first term above is non-vanishing in the situation where the rotor decays.

\para
Our main interest lies in the quantum numbers of the outgoing fermions, and how they are affected by the rotor. To proceed, the expression \eqn{eom-in-terms-of-Jtilde} will prove most useful. In the low-energy limit  $E\ll 1/I$, we may drop the inertia term in \eqn{eom-in-terms-of-Jtilde}, leaving
\be \dot{\alpha}(t) = \frac{4\pi}{N} \tilde{J}_{\rm out}(t)\
\label{eq:rotoreom2}
.\ee
Integrating this once  gives
\be 
\alpha(t-x) 
= \alpha(t) - \frac{4\pi}{N}\int_{t-x}^t dt'\, \tilde{J}_{\rm out}(t')
\,.\label{alphaj}
\ee
The current $\tilde{J}_{\rm out}(t')$ is defined in terms of $\psi_i(t,0^+)$, the fermions just to the right of the rotor in the $r_0 \ra 0$ limit.
Because the fermions are just free massless right-movers away from the rotor, we have $\psi_i(t,0^+) = \psi_i(t+T, T)$ for any $T>0$.
Similarly, $\psi_i(t,0^-) = \psi_i(t-T, -T)$ for any $T>0$. 
Then, for any fermion $\psi(t,y)$ and the rotor $\alpha(t-x)$ with $x>0$, 
we have from \eqn{alphaj} that 
\be 
[\alpha(t-x), \psi_i(t,y)] 
&=& -\frac{4\pi}{N}\int_{t-x}^t  dt'\, [\tilde{J}_{\rm out}(t'), \psi_i(t,y)] \nn\\ 
&=& -\frac{4\pi}{N} \int_{t-x}^{t} dt'\, \sum_j[\psi_j^\dag \psi_j(t', 0^+), \psi_i(t, y)] \nn\\ 
&=& -\frac{4\pi}{N} \int_{t-x}^{t} dt'\, \sum_j[\psi_j^\dag \psi_j(T, -t'+T), \psi_i(T, y-t+T)] \nn\\ 
&=& -\frac{4\pi}{N} \int_{t-x}^t dt'\, \delta(t'-t+y) \,\psi_i(t,y) 
\,.
\ee
Here, for $y<0$, the shifts are justified because there exists $T$ such that $-t'+T > 0$ and $y-t+T<0$ over the entire integration range $t-x < t' < t$.
However, the $\delta$-function in the last line is never picked up in this integration range, so the commutator vanishes.
For $y>0$ (and still $x>0$), the shifts are again justified because there exists $T$ such that $-t'+T > 0$ and $y-t+T > 0$, 
and this time the $\delta$-function does get picked up as long as $-x+y < 0$.
We thus have
\be 
[\alpha(t-x), \psi_i(t,y)]
= -\frac{4\pi}{N} \, \theta(y) \, \theta(x-y) \, \psi_i(t,y)
\,.\label{rotorpsicomm}
\ee

\begin{wrapfigure}{r}{0.44\textwidth}
\centering
\includegraphics[width=.7\linewidth]{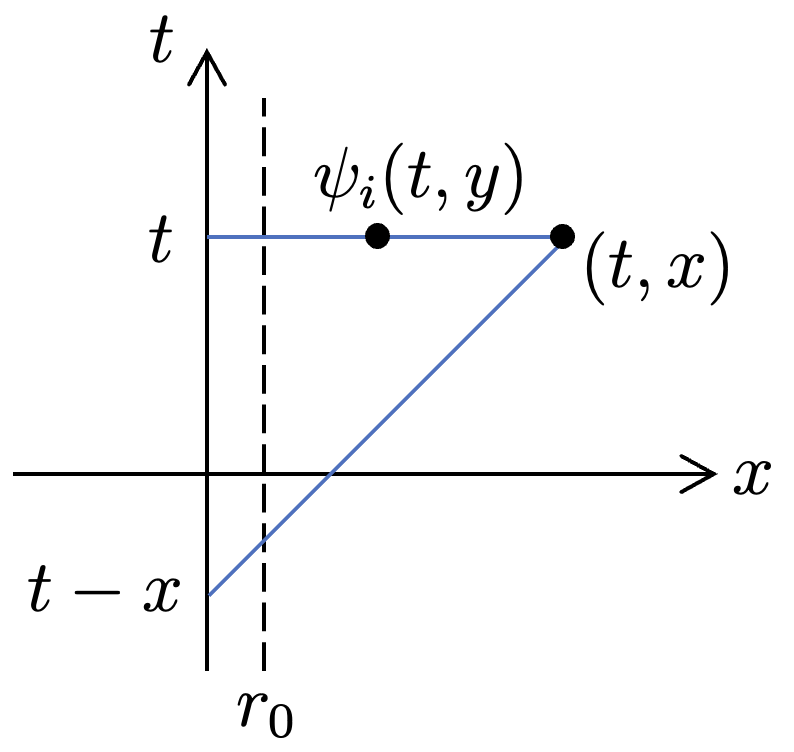}
\caption{The space-time diagram of fermion-rotor insertion.}
\end{wrapfigure}
\noindent
The product of step functions $\theta(y)\theta(x-y)$ is telling us that the rotor and fermion fail to commute only if $0<y<x$, so that fermion insertion $\psi_i(t,y)$ lies within the lightcone of the rotor interaction at time $t-x$, as shown in the figure to the right. For $y>x$ or $y<0$, the rotor and fermion commute.

\para
We define the twist operator ${\cal T}(t,x)$, for $x>0$ only, as
\be {\cal T}(t,x) = e^{i\alpha(t-x)}\ \ .
\label{eq:twist}
\ee
From \eqn{rotorpsicomm}, this commutes with ingoing fermions, $\psi_i(t,y)$ for $y<0$, but for outgoing fermions the operator enjoys the equal-time exchange relations
\be {\cal T}(t,x) \psi_i(t,y) = \psi_i(t,y) {\cal T}(t,x)\ \ \ &&\mbox{for\ $y>x$}\nn\\
{\cal T}(t,x) \psi_i(t,y) = e^{-4\pi i/N} \psi_i(t,y) {\cal T}(t,x)\ \ \ &&\mbox{for\ $y<x$}\ .\label{twistcom}\ee
These are indeed the exchange relations of a twist operator (see, for example, \cite{cardy}). In the present case, the twist operator acts by implementing a $e^{-4\pi i/N}$  phase rotation  on outgoing fermions. This agrees with the proposal of \cite{me}, where the same twist was derived using conformal field theory techniques\footnote{Note that our twist operator here is the inverse-square of the twist in \cite{me}: ${\cal T}_{\rm here} = T^{-2}_{\rm there}$.}. 

\para
There is, however, a difference between the BCFT techniques used in \cite{me} and the rotor model. In the low-energy conformal limit, the twist operator is attatched to a topological line, which can end anywhere on the monopole worldline. This is a consequence of taking the low-energy limit. In contrast, for us it's crucial that the attached rotor degree of freedom is evaluated at the retarded time $t-x$, when the fermion passed through the origin. A similar phenomenon is seen in lattice models that exhibit similar properties \cite{ueda}.

\para
We can further make contact with \cite{me} if we bosonize the fermions. Sweeping a UV cut-off under the rug, the bosonisation dictionary is 
\be \psi_i(t,x<-r_0)=\frac{1}{\sqrt{2\pi}}e^{i\chi_i(t-x)} \ \ \ {\rm and}\ \ \ \psi_i(t,x>r_0)=\frac{1}{\sqrt{2\pi}}e^{i\tilde{\chi}_i(t-x)}\ .\label{boson}\ee
The outgoing current is then given by $\tilde{J}_{\rm out} = -\sum_i\partial_+\tilde{\chi}_i/4\pi$. We can then integrate \eqn{eq:rotoreom2} to get an expression for the rotor, and hence the twist operator, directly in terms of the bosonised fields
\be {\cal T}(t,x) = \exp\left(-\frac{2i}{N}\sum_i \tilde{\chi}_i(t-x)\right)\ .
\label{eq:twistN}
\ee
This too agrees with the results of \cite{me}, and gives a straightforward way to extract the physics. For example, in the case of $N=2$, we have ${\cal T} = e^{-i(\tilde{\chi}_1+ \tilde{\chi}_2)}$. In this case, propagation through the rotor gives
\be \psi_1 \sim e^{i\chi_1} \rightarrow  {\cal T} e^{i\tilde{\chi}_1} = e^{-i\tilde{\chi}_2} \sim \psi_2^\dagger
\label{eq:twistN2}
\ee
The propagation of $\psi_1$ and $\psi_2$ is given by  $\psi_i\ra \epsilon_{ij}\psi_j^\dagger$.  

%

\subsection{More Rotors, More Twists}\label{chargesec}

In this section, we look at a more general class of fermion-rotor models. We will again consider $N$ fermions, $\psi_i(t,x)$, with $i=1,\ldots, N$, but now we take these to interact with $r$ different rotor degrees of freedom, $\alpha_a(t)$, with $a=1,\ldots,r$. Each of these rotors is  localised at some point, but they need not be at the same point. We introduce a collection of regularised delta-functions $f'_a(x)$, each obeying
\be \int_{-\infty}^{\infty} dx\ f'_a(x)  = 1\ .\ee
Moreover, we allow each fermion to interact with  the rotors through different charges $q_{ai}\in\Z$.  The action of this model is 
\be S = \int d^2x \ i\sum_{i=1}^N\psi_i^\dagger\big(\partial_++ i\sum_{a=1}^rq_{ai} \alpha_a(t) f'_a(x)\big) \psi_i + \ \sum_{a=1}^r\int dt\ \frac{I_a}{2} \dot{\alpha_a}^2\ .\label{morecharge}\ee
As in the previous section, we can think of this as $N$ chiral fermions, coupled to $r$ gauge fields. For a general choice of $f_a(x)$, with rotors localised at different points, the theory is consistent, with no mixed gauge anomalies, providing that the charges obey
\be \sum_{i=1}^N q_{ai} q_{bi} = 0 \ \ {\rm for} \ a\neq b.\ee
(We can relax this condition when some of the $f_a(x)$ coincide, although we will choose not to here.) There is a global symmetry associated to each rotor, given by
\be \psi_i(t,x) \ra e^{iq_{ai}\beta_a\,f_a(x)}\psi_i(t,x)\ \ \ {\rm and}\ \ \ \alpha_a(t) \ra \alpha_a(t) - \beta_a\ \  \ {\rm with}\ a=1,\ldots, r\ .\ \ \ \  \label{symf2}\ee
These are the extensions of the previous symmetry \eqn{symf}. In addition, there are further global symmetries that leave the rotors untouched. To describe these, we first introduce a collection of charges $p_{a'i}\in\Z$ with $a'=r+1,\cdots, N$, satisfying
\be \sum_{i=1}^N q_{ai}\,p_{a'i} = 0\ .\label{qcond}\ee
This orthogonality condition ensures that the following  $U(1)^{N-r}$ symmetries labelled by $a'$ are  non-anomalous
\be \psi_i(t,x) \ra e^{ip_{a'i}\beta_{a'}} \psi_i(t,x)\ \ \ {\rm with}\  a'=r+1,\ldots, N\ .\ee
The upshot is that there are $U(1)^N$ global symmetries, that naturally decompose as $U(1)^r \times U(1)^{N-r}$. The factors of $f_a(x)$ in \eqn{symf2} mean that the $U(1)^r$ symmetries act differently on the incoming and outgoing fermions (and, indeed, differently yet again on the fermions in regions between rotors). We denote the charges of fermions on the far left as $Q_{\alpha i}$ and the charges of fermions on the far right as $\bar{Q}_{\alpha i}$, with $i=1,\ldots,N$ labelling fermions and $\alpha= (a,a')=1,\ldots, N$ labelling symmetries. These are given by
\be Q_{\alpha i} = \left(\begin{array}{rr}  -q_{ai} \\ p_{a'i}\end{array}\right)\ \ \ {\rm and}\ \ \ \bar{Q}_{\alpha i} =  \left(\begin{array}{cc}  q_{ai} \\ p_{a'i}\end{array}\right)\ .\label{ccbar}\ee
%
By construction, these obey $Q_{\alpha i}Q_{\beta i} = \bar{Q}_{\alpha i}\bar{Q}_{\beta i}$ which is the condition for anomaly cancellation.

As an example, consider $N=2$ fermions coupled to a single rotor with charges $q_i$. The charges for fermions to the left and right of the rotor are
\be Q = \left(\begin{array}{cc} -q_1\ & -q_2 \\ q_2\ & -q_1\end{array}\right)\  \ \ \ {\rm and}\ \ \ \bar{Q} = \left(\begin{array}{cc} q_1\ & q_2 \\ q_2\ & -q_1\end{array}\right)\ .\ee
This simple set-up includes the well-studied 3450 model, in which the two incoming fermions carry charge 5 and 0 under a $U(1)$ symmetry, while the two outgoing fermions carry charge 3 and 4. This arises, for example, if we take  $q_i=(1,2)$. In this case, the 3450 symmetry arises from the linear combination $- Q_{1i} +2Q_{2i} = (5,0)$ and $- \bar{Q}_{1i} + 2\bar{Q}_{2i} = (3,-4)$, 

\para
In the context of BCFT, the boundary state arising from left and right-moving fermions carrying charges specified by charges $Q$ and $\bar{Q}$ respectively were discussed in detail in \cite{philip1,philip2}. (Some results were found previously in  \cite{costas}.) There it was shown that the right-moving and left-moving currents are related by the rotation matrix
\be
\mathcal{R}_{ij}= (\bar{Q}^{-1})_{i\alpha}Q_{\alpha j}=\delta_{ij}-2\sum_a \frac{q_{ai} q_{aj}}{\sum_k q_{ak}^2}
\ .\label{rij}
\ee
Here we demonstrate that the same result arises in the low-energy limit of the rotors.

\para
To connect with the  BCFT result we determine the boundary condition implemented by rotors which, for simplicity, we assume are all located at the origin (although we still require \eqn{qcond}).  Solving for the equation of motion near the core gives
\be  \psi_i(t,-r_0)=\psi_i(t,r_0)e^{i \sum_a q_{ai} \alpha_a(t)}\ .
\label{eq:boundarycondition}
\ee
It is convenient to perform Abelian bosonization  \eqn{boson}.  Then, repeating the steps that lead to \eqn{eq:rotoreom2}, we have, at low energies, 
\be \dot{\alpha}_a(t) \approx \frac{4\pi}{\sum_i q_{ai}^2} \sum_i q_{ai} \tilde{J}_{{\rm out},i}(t)=-\frac{2}{\sum_i q_{ai}^2} \sum_i q_{ai} \partial_t\tilde{\chi}_i(t)\label{adotmore}
\ee
where we've introduced the outgoing current
\be \tilde{J}_{{\rm out},i}  = \lim_{r_0\ra0} \psi^\dagger_i\psi_i (t+r_0,r_0) = -\frac{1}{2\pi} \,\partial_t\tilde{\chi}_i(t) \ .\ee
Integrating the equation of motion \eqn{adotmore} and substituting into the boundary condition  \eqn{eq:boundarycondition} we get, up to an arbitrary phase, the matching condition across the rotor
\be  e^{i\chi_i} = e^{i\sum_j {\cal R}_{ij} \tilde{\chi}_j}
\label{chirij}\ee
with the rotation matrix ${\cal R}_{ij}$ given by \eqn{rij}. This same rotation matrix can be used to match currents across the rotor, in agreement with the BCFT results.

\para
We define a  twist operator ${\cal T}_a(t,x)$, for $x>0$, for each rotor
\be {\cal T}_a(t,x) = e^{i\alpha_a(t-x)}\ \ .
\ee
Again, this commutes with ingoing fermions, $\psi_i(t,y)$ for $y<0$. For outgoing fermions, with $y>0$, the  generalisation of \eqn{twistcom} is the equal-time exchange relations
\be {\cal T}_a(t,x) \psi_i(t,y) = \psi_i(t,y) {\cal T}_a(t,x)\ \ \ &&\mbox{for\ $y>x$}\nn\\
{\cal T}_a(t,x) \psi_i(t,y) = \exp\left(-\frac{4\pi i q_{ai}}{\sum_k q_{ak}^2} \right) \psi_i(t,y) {\cal T}_a(t,x)\ \ \ &&\mbox{for\ $y<x$}\ .\ee
For example, in the 3450 model, which consists of a single rotor with charges $q_i=(1,2)$, the twist operator obeys, for $0<y<x$,
\be {\cal T}(t,x) \psi_1(t,y) &=& e^{-4\pi i/5} \psi_1(t,y) {\cal T}(t,x)\nn\\ {\rm and}\ \ \ \ {\cal T}(t,x) \psi_2(t,y) &=& e^{-8\pi i/5} \psi_2(t,y) {\cal T}(t,x)\ .\ee
These are the twists identified in \cite{me}.

\para
Again, there is free propagation across the rotor. This time, an incoming fermion $\psi_i$ will evolve to an outgoing fermion $(\prod_a {\cal T}_a^{q_{ai}})\psi_i$.

\section{Correlation Functions}\label{corrsec}

In this section, we calculate correlation functions in the more general class of fermion-rotor models \eqn{morecharge}. To keep things simple, we restrict to just a single rotor $\alpha(t)$, localised at the origin, coupled to $N$ fermions with charges $q_i$ (which is a shorthand for saying they have charges $-q_i/2$ and $q_i/2$ at $x<-r_0$ and $x>r_0$, respectively). The extension to multiple rotors is straightforward.  We work with the action
\be S = \int d^2x \ i\sum_{i=1}^N \psi_i^\dagger \left(\partial_++ iq_{i} \alpha(t) f'(x)\right)\psi_i  + \int dt\ \frac{I}{2} \dot{\alpha}^2\ .\label{oneqact}\ee
We evaluate the Lorentzian partition function  of this theory in Appendix \ref{appa}, where we show that, at low energies $E\ll 1/I$,  fermionic correlation functions take the form
\be \Big< \prod_{j=1}^n\psi_{i_j}(t_j,x_j)\prod_{k=1}^{n'}\psi^\dagger_{i'_k}(t'_k,x'_k)\Big> = (\mbox{free correlators})\times {\cal Z}\label{fcorr}\ee
with 
\be {\cal Z} = \exp\left(\frac{2}{\sum_i q_i^2} \int_0^\infty \frac{d\omega}{\omega} AB\right)\ .\label{zab}\ee
Here the two functions $A$ and $B$ are given by
\be A &=& \sum_{j=1}^n q_{i_j}\theta(x_j) e^{-i\omega (t_j-x_j)} -  \sum_{k=1}^{n'} q_{i'_k}\theta(x'_k) e^{-i\omega (t'_k-x'_k)}\nn\\
 B &=& \sum_{j=1}^n q_{i_j}\theta(-x_j) e^{+i\omega (t_j-x_j)} -  \sum_{k=1}^{n'} q_{i'_k}\theta(-x'_k) e^{+i\omega (t'_k-x'_k)}
 \ .\label{aandb}\ee
 The function  $A$ gets contributions from insertions to the right of the rotor, and the function $B$ from insertions to the left.  We will also want to compute correlation functions with rotor insertions. Typically, we will want these insertions to take place at either  $x>0$ or $x<0$. In the former case, we write
 \be e^{iq\alpha(\tau)\theta(x) } = \exp\left(iq\int_0^\infty \frac{d\omega}{2\pi} \theta(x) \left[\alpha(\omega)e^{-i\omega\tau} + \alpha(-\omega) e^{i\omega \tau}\right]\right)\ .\ee
Following the derivation in the appendix (see, in particular, the discussion around \eqn{a11}),  we can accommodate a rotor insertion $e^{iq\alpha(\tau)\theta(x)}$ at $x>0$ by shifting
 \be A \ra A - q\theta(x) e^{-i\omega \tau} \ \ \ {\rm and}\ \ \ B\ra B +  q\theta(x) e^{i\omega \tau}\ .
 \label{eq:rotorshift1}
 \ee
 Similarly, an insertion of $e^{-iq\alpha(\tau)\theta(-x)}$ at $x<0$ is implemented by the shift
 \be A \ra A + q\theta(-x) e^{-i\omega \tau} \ \ \ {\rm and}\ \ \ B\ra B -  q\theta(-x) e^{i\omega \tau}\ .\ee
Either with or without rotors, we're left with the integral \eqn{zab}.
The integral is UV convergent owing to rapid oscillations of the form $e^{-i\omega s}$ at large $\omega \gg 1/|s|$, 
while it may be divergent in the IR\@.
We regularise with an IR cutoff $\mu > 0$ as
 \be \int_0^\infty \frac{d\omega}{\omega} \, e^{-i\omega s} 
 \longrightarrow \int_\mu^\infty \frac{d\omega}{\omega} \, e^{-i\omega s}=  - \gamma - \log(i\mu s) + \ldots
 \label{intreg}\ee
 where the $\ldots$ represent terms that vanish as $\mu \ra 0$, and $s$ should be understood as $s-i0^+$ to deal with the branch cut of the logarithm.   This regularisation is not without issue; we will return to this in Section \ref{clustersec}.

\para
We can now use this prescription to compute some simple correlation functions. We start with two-point functions. The result \eqn{fcorr} is proportional to the free correlator,  which means that $\langle \psi_i(t,x)\psi_j(t',x')\rangle=0$. (There is actually a subtlety here that we discuss further in  Section \ref{clustersec}.) To get a non-vanishing answer, we should look at $ \langle \psi_i(t,x)\psi^\dagger_j(t',x')\rangle$, which will be proportional to $\delta_{ij}$.

\para
First,  as a sanity check to make sure that the general formalism is sensible, if  $xx'>0$, so that fermions are created and annihilated without passing through the rotor, we get either $A=0$ or $B=0$  and so the integral \eqn{zab} is just ${\cal Z}=1$ and
\be\langle \psi_i(t,x)\psi^\dagger_j(t',x')\rangle=\delta_{ij}G_0(t-t',x-x') \ .\ee
with 
\be G_0(t,x) = \frac{1}{2\pi i}\frac{t+x}{t^2-x^2-i0^+}\label{freeprop}\ee
the propagator for a free chiral fermion. This is the expected result.

\para
Things are more interesting when fermions are created and annihilated on different sides of the rotor. If  $x>0$ and $x'<0$ and then
\be A =  q_{i}e^{-i\omega (t-x)} \ \ \ {\rm and}\ \ \  
B =-q_{j}e^{i\omega (t'-x')} \ .\label{ABfermion}\ee
From \eqn{zab}, the IR-regulated integral gives ${\cal Z}$ a contribution that scales as $\mu^{q_iq_j}$ which, for $i=j$ where the free correlators are non-zero, vanishes as $\mu\ra 0$. Thus, 
\be \langle \psi_i(t,x)\psi^\dagger_j(t',x')\rangle=0 \ .
\label{psi-psidag-opposite-sides}\ee
The correlator vanishes also if $x<0$ and $x'>0$. This means that if a fermion is created one side of the rotor, and annihilated on the other side, then the correlation function vanishes \cite{joe}. In some sense, this result is unsurprising because the vanishing is imposed by the $U(1)$ global symmetry \eqn{symf}. Nonetheless, it does lay bare the general scattering paradox: if you create a fermion to the left of the rotor, what comes out the other side?

\para
The key observation of \cite{lt} is that a non-vanishing two-point function arises for $xx'<0$ if we insert the rotor to balance $U(1)$ charge at exactly the right moment. For example, we can consider $x>0$, $x'<0$ and evaluate
\be\langle \psi_i(t,x)e^{iq_i\alpha(t-x)}\psi^\dagger_j(t',x')\rangle\ .\ee
It is important that the rotor is evaluated at the time $t-x$ at which the fermion passed through the origin. 
As shown in \eqn{eq:rotorshift1}, the rotor insertion shifts $A$ and $B$ compared to  \eqn{ABfermion} as $A\to A-q_ie^{-i\omega(t-x)}=0$ and $B\to B+q_ie^{i\omega(t-x)}$. This results in the free propagator
\be\langle \psi_i(t,x)e^{iq_i\alpha(t-x)}\psi^\dagger_j(t',x')\rangle=\delta_{ij}G_0(t-t',x-x') .\label{2pac}\ee
We see that dressing with the rotor, which, as shown in Section \ref{frsec}, acts a twist operator, the fermion propagates freely through the impurity. 

\para
We could also ask what happens if we evaluate the rotor at a time different from $\tau=t-x$. In this case, the integral \eqn{zab} has a UV-divergence. This can be traced to the fact that we're working in the low-energy limit $E\ll 1/I$. If we identify the UV cut-off with $1/I$, then the integral scales as $I^{2/N}$, reflecting the fact that such a correlation function requires the on-shell excitation of the rotor. 
\para
In general, we can define the operator
\be
\mathcal{O}_i(t,x) 
=
\e^{iq_i\alpha(t-x) \theta(x)} \psi_i(t,x)
\ .\ee
This operator transforms as $\mathcal{O}_i(t,x)\to \mathcal{O}_i(t,x)e^{-iq_i\beta/2}$ under the $U(1)$ symmetry \eqn{symf}. The operator  looks non-local: the fermion operator $\psi_i(t,x)$ is dressed with the rotor evaluated at the time $t-x$ in the past when the two interacted. And yet, as we've seen, the non-locality is mild. Indeed, the operator $\mathcal{O}(t,x)$ is, in some sense, free, 
with two-point function $\langle {\cal O}_i(t,x)\,{\cal O}^\dagger_j(t',x')\rangle = \delta_{ij} G_0(t-t',x-x')$ for any $x$ and $x'$. However, higher point correlators exhibit branch cuts that capture the mild non-locality inherent in twist operators \cite{lt}.

\para
There is a straightforward way to create the seemingly ``non-local'' state ${\cal O}_i^\dagger(t,x)\ket{0}$ with $x>0$: just create the state ${\cal O}_i^\dagger(t-x,x)\ket{0} = \psi^\dagger_i(t-x,x)\ket{0}$ for $x<0$ and wait. In the language of \cite{fock,me,me2}, these are the states in the twisted Fock space. 

\para
We can also introduce the related operator 
\be 
\tilde{\mathcal{O}}_i(t,x) 
=
\e^{-iq_i\alpha(t-x) \theta(-x)} \psi_i(t,x) 
\ee
which transforms as $\tilde{\mathcal{O}}_i(t,x)\to \tilde{\mathcal{O}}_i(t,x)e^{+iq_i\beta/2}$ under the $U(1)$ symmetry \eqn{symf}. This too obeys $\langle \tilde{\cal O}^\dagger_i(t,x)\,\tilde{\cal O}_j(t',x')\rangle = \delta_{ij} G_0(t-t',x-x')$. Now, note that the operator is inserted for fermions $\psi(t,x)$ inserted to the left of the rotor, and is evaluated at the time $t-x$ in the future (because $x<0$). While we can certainly write down  this operator, in contrast to ${\cal O}(t,x)$, it is difficult to see how we can construct the state $\tilde{\cal O}(t,x)\ket{0}$, with $x<0$,  by a local experiment.

\subsection{Cluster Decomposition}\label{clustersec}

Things are more subtle than we've so far made out. We can highlight this by considering the case of a single $N=1$ fermion with charge $q=1$ interacting with the rotor. Here there's no ambiguity in identifying the final state: the symmetry \eqn{symf} means that, up to an overall phase,  an ingoing fermion must scatter to 
\be \psi \ra \psi^\dagger\ .\ee
This equation only makes sense if $\psi$ at $x<0$ scatters to $\psi^\dagger$ at $x>0$; only then is the $U(1)$ symmetry \eqn{symf} obeyed. Namely, this $\psi^\dagger$ has the same $U(1)$ charge as the $\psi$, and as such it does not refer to the antiparticle of the $\psi$. However, the ${}^\dagger$ does indicate that the final state carries an opposite fermion number to the initial state, which is an anomalous symmetry  \eqn{notsym} due to the rotor so the fermion number is not conserved as the fermion pass through the rotor.

\para
If we were to fold the system, so the rotor acts as a boundary rather than an impurity, then this is Andreev reflection, in which an electron reflects off a superconductor and returns as a hole.  In this context, the physical electric charge is identified with the symmetry \eqn{notsym}, which is anomalous in our model, but is spontaneously broken in a superconductor.

\para
This scattering process means that the two-point function $\langle \psi(t,x)\psi(t',x')\rangle$ across the rotor should be non-zero. Yet the calculations described above suggests that it vanishes. Clearly something is afoot.

\para
In fact, the calculation of the two-point correlator  $\langle \psi(t,x)\psi(t',x')\rangle$ has two factors, as shown in \eqn{fcorr}. The first is the free propagator which manifestly vanishes. The second is the integral and this has an IR divergence as we remove the regulator. So the answer is $0\times \infty$. What to do?

\para
We can get some insight into this by noting that a non-vanishing correlator $\langle \psi\psi\rangle$ violates the anomalous fermion number symmetry \eqn{notsym}. The anomaly equation \eqn{anomaly} means that this must come about from an instanton-like configuration in which the rotor winds, with the change in fermion number $\Delta Q_\psi$ given by
\be \Delta Q_\psi  = \frac{N}{2\pi}\big(\alpha(t\ra +\infty) - \alpha(t\ra -\infty)\big)\ .\ee
Scattering processes of this form, that violate the anomalous symmetry but do not, otherwise, involve anything strange like twisted sectors, are analogous to the Callan-Rubakov effect for fermion-monopole scattering that results in proton decay \cite{c,r}.
The method that we described above works for correlators with $\Delta Q_\psi=0$. 
As noted in \cite{joe}, the reason that it does not work for $\Delta Q_\psi \neq 0$ is our IR cutoff. Since this removes low frequency modes with $\omega < \mu$ in $\alpha(t)$, it is not possible for $\alpha(t)$ to asymptote to different values as $t \ra \pm\infty$. Then, the anomaly relation above restricts us to consider only $\Delta Q_\psi = 0$.
Here we want to generalise to correlators with $\Delta Q_\psi \neq 0$.

\para
A simple way to proceed was suggested by Polchinski in \cite{joe}. Consider an arbitrary operator $\Theta(t_1,\ldots,t_n,x_1,\ldots,x_n)$ depending on $n$ spacetime positions. We want to calculate $\langle \Theta\rangle$. If this operator carries non-zero charge under the anomalous symmetry then the method used above is ambiguous: $0\times \infty$. Instead we compute the correlator
\be
\left\langle \Theta(t_1,...,t_n,x_1,...,x_n)\,\Theta^\dagger(t'_1,...,t'_n,x'_1,...,x'_n)\right\rangle \,,
\ee
This has vanishing anomalous charge and can be calculated faithfully using the methods above. We then take the limit $t_i\gg t'_j$. Cluster decomposition requires that 
\be
\left\langle \Theta(t_1,...,t_n,x_1,...,x_n)\,\Theta^\dagger(t'_1,...,t'_n,x'_1,...,x'_n)\right\rangle &\longrightarrow&
\big\langle \Theta(t_1,...,t_n,x_1,...,x_n)\big\rangle \\ &&\ \ \ \ \ \ \times\ \big\langle\Theta^\dagger(t'_1,...,t'_n,x'_1,...,x'_n)\big\rangle \,.
\nn\ee
We now apply this cluster decomposition method to the $N=1$ case of Andreev scattering. We'll then see what light it has to shed on more general examples, and the role of the twist operator.

\para
For a single $N=1$ fermion, we want to calculate $\langle \psi(t_1,x_1)\psi(t_2,x_2)\rangle$ with $x_1<0$ and $x_2>0$.  This violates the anomalous fermion number symmetry with $\Delta Q_\psi=2$. It is, however, straightforward to compute the four-point function
\be \Gamma_{N=1}(t_i,x_i;t'_i,x'_i) = \big\langle\psi(t_1,x_1)\psi(t_2,x_2)\left(\psi(t'_1,x'_1)\psi(t'_2,x'_2)\right)^\dagger\big\rangle \ee
using the methods of the previous section. For $x_1,x'_1<0$ and $x_2,x'_2>0$, it is given by 
\be
\Gamma_{N=1}(t_i,x_i;t'_i,x'_i) =  \big( G_{11'} G_{22'} - G_{12'}G_{21'} \big)\times \left(  \frac{s_{12'}s_{1'2}}{s_{12}s_{1'2'}} \right)^2  
\ .\ \ \ \ \label{n1cluster}\ee
Here we define the lightcone coordinate $s_i = t_i-x_i$ and $s_{ij} = s_i - s_{j}$. The first factor in \eqn{n1cluster} contains the free propagators
\be G_{ij'} = G_0(t_i-t'_j,x_i-x_j') = \frac{1}{2\pi i} \frac{1}{s_{ij'}}\  .\ee
The second factor in \eqn{n1cluster} comes from the integration in \eqn{fcorr}. Now we seperate the clusters by a large time $T$, so that $t_i-t'_i\sim T$. The contribution from the integration scales as $T^4$. Each of the propagators scales as $G_{ij'}\sim 1/T$, but the fermionic Wick contractions, resulting in the minus sign in the first bracket in \eqn{n1cluster}, ensures that the combination of propagators scales as $1/T^4$. The upshot is that, in the limit $T\ra \infty$, we have
\be \Gamma_{N=1}(t_i,x_;t'_i,x'_i) \longrightarrow  -\left(\frac{1}{2\pi i}\right)^2 \frac{1}{s_{12}}\frac{1}{s_{1'2'}}\ .\ee
This exhibits the desired cluster decomposition. We learn that the   two-point correlator is given by 
\be \langle \psi(t_1,x_1)\psi(t_2,x_2)\rangle = \frac{e^{i\theta}}{2\pi i} \frac{1}{s_{12}}\ .\ee
Here $e^{i\theta}$ is some overall phase that is undetermined by the analysis above. (Indeed, the possible boundary conditions that implement Andreev reflection also allow for such a phase.)

\para
There is a similar story when we look at more general models, but now with an additional ingredient.  To see this, consider the case of  $N=2$ fermions interacting with a single rotor with charges $q_i=(1,1)$. In this case there is an $SU(2) \times U(1)$ symmetry; incoming fermions carry charges ${\bf 2}_{-1}$ and outgoing fermions ${\bf 2}_{+1}$. We start by studying the outgoing fermion attached to a rotor degree of freedom, 
\be \psi_i \ra e^{i\alpha} \psi_i\ .\ee
Unlike for $N>2$, there is no fractional power in the exponent $e^{i\alpha}$, ensuring that, for $N=2$, this does not involve a twist. This scattering is confirmed by the two-point correlator \eqn{2pac}. 
Following our analyses of the twist operator in the theory with $N=2$ \eqn{eq:twistN2}, we have an operator relation

\be e^{i\alpha(t-x)}\psi_i(t,x) = \epsilon_{ij}\psi_j^\dagger(t,x)\ ,\label{same}\ee
hence, there is no twisted mystery for this particular scattering process, and scattering corresponds to the well-known Callan-Rubakov process
\be \psi_i \ra \epsilon_{ij} \psi^\dagger_j\ .
\label{eq:N2scat}
\ee
This is possible only for $N=2$ because the representation ${\bf 2}$ is pseudoreal. In general, an operator containing $N/2$ coinciding rotor insertions can be represented as an excitation in  the naive Fock space. This can be easily seen by noting that the twist operator \eqn{eq:twistN} is raised to  the power of $N/2$, so $\mathcal{T}^{N/2}$, represents a well-defined vertex operator that can be decomposed into fermionic excitations in the naive Fock space.

\para
%

%
\para
It is straightforward to use cluster decomposition to confirm the operator relation \eqn{eq:N2scat} when inserted in the two-point function $\langle \psi_i(t_i,x_j)\psi_j(t_j,x_i)\rangle$, and show that it is proportional to $\epsilon_{ij}$ multiplied by the free propagator. Indeed, these are calculations previously performed in \cite{joe}.  Specifically, we will calculate $\langle\psi_2(t_2,x_2)\psi_1(t_1,x_1)\rangle$. As in the previous example, we do this by instead looking at the four-point function
\be \Gamma_{N=2}(t_i,x_i;t'_i,x'_i) = \big\langle\psi_2(t_2,x_2)\psi_1(t_1,x_1)\left(\psi_2(t'_2,x'_2)\psi_1(t'_1,x'_1)\right)^\dagger\big\rangle \ee
which is easily calculated using the methods using the general result \eqn{fcorr}. For $x_1,x'_1<0$ and $x_2,x'_2>0$, it is given by 
\be
\Gamma_{N=2}(t_i,x_i;t'_i,x'_i) =  G_{11'}G_{22'} \frac{s_{21'}s_{2'1}}{s_{21}s_{2'1'}}
\ee
In the limit $t_i\gg t'_j$, the correlation function reduces to
\be
\Gamma_{N=2}(t_i,x_i;t'_i,x'_i) &\longrightarrow& \big\langle\psi_2(t_2,x_2)\psi_1(t_1,x_1)\big\rangle
\big\langle\left(\psi_2(t'_2,x'_2)\psi_1(t'_1,x'_1)\right)^\dagger\big\rangle \nn\\ &=&  -\frac{1}{(2\pi i)^2}\frac{1}{s_{21}s_{2'1'}} \,.
\ee
This allows us to read off  two separate correlators, up to an overall phase
\be
\big\langle\psi_2(t_2,x_2)\psi_1(t_1,x_1)\big\rangle =\frac{e^{i\theta}}{2\pi i}\frac{1}{s_{21}} \label{freeagain}\ee
and 
\be
\big\langle\psi^\dagger_1(t'_1,x'_1)\psi^\dagger_2(t'_2,x'_2)\big\rangle =-\frac{e^{-i\theta}}{2\pi i}\frac{1}{s_{2'1'}} \,.
\ee
Again, we have the possibility of an overall phase $e^{i\theta}$. (For a discussion of such phases in the context of boundary conditions for chiral fermions, see \cite{philip1}.)   The correlation function \eqn{freeagain} is, again, that of a free fermion. 

\para
It is also interesting to compute a similar four-point function
\be \Gamma'_{N=2}(t_i,x_i;t'_i,x'_i) = 
 \big\langle\psi_2(t_2,x_2)e^{i\alpha(t_1-x_1)}\psi_1(t_1,x_1)\left(\psi_2(t'_2,x'_2)e^{i\alpha(t'_1-x'_1)}\psi_1(t'_1,x'_1)\right)^\dagger\big\rangle \ .\nn\ee
This time all $x_{1,2}, x'_{1,2} > 0$.
The calculation proceeds nearly identically as the one for $\Gamma_{N=2}$
and we obtain 
\be
\big\langle\psi_2(t_2,x_2)e^{i\alpha(t_1-x_1)}\psi_1(t_1,x_1)\big\rangle =\frac{e^{i\theta'}}{2\pi i}\frac{1}{s_{21}} \ee
with some $\theta'$. As discussed in \cite{lt}, this shows that the operators $\psi_2$ and $e^{i\alpha(t-x)}\psi_1(t,x)$ interpolate the same one-particle state. 
We can further compute
\be\big\langle\psi_2(t_2,x_2)e^{i\alpha(t_1-x_1)}\psi_1(t_1,x_1)\left(\psi_2(t'_2,x'_2)\psi_1(t'_1,x'_1)\right)^\dagger\big\rangle\ee
at $x_{1,2}, x_2' > 0$ and $x'_1 <0$ 
to show that $\theta' = \theta$, 
which thus provides a confirmation of the operator relation~\eqn{same} directly in the fermion-rotor framework.
It is interesting that this operator relation is much more manifest in the bosonic language \eqn{eq:twistN2}.


\para
As our final example, we look at the 3450 model which, as we saw in Section \ref{chargesec} is described by a theory with $N=2$ and charges $q_i=(1,2)$. In this case, the Callan-Rubakov scattering  which do  not involve any twist operators are the $3\ra 3$ processes
\be 2\psi_1 + \psi_2^\dagger \ra 2\psi_1 + \psi_2^\dagger\ \ \ {\rm and}\ \ \ \psi_1^\dagger + 2 \psi_2^\dagger \ra \psi_1 + 2\psi_2 \label{twoprocess}\ee
Here we compute the six-point correlation function associated to each of these scattering processes. The first process is clearly consistent with three separate free propagations but, as we now see, the correlation function is somewhat more complicated. We consider the 6-point function
\be \Theta_{11\bar{2}}(t_i,x_i) = \psi_1(t_1,x_1)\psi_1(t_2,x_2)\psi_2^\dagger(t_3,x_3)\psi_1^\dagger(t_4,x_4)\psi_1^\dagger(t_5,x_5)\psi_2(t_6,x_6)\ .\ee
We start by computing the 12-point function
\be \langle \Theta_{11\bar{2}}(t_i,x_i)\Theta_{11\bar{2}}^\dagger(t'_i,x'_i)\rangle\ee
where we take $x_{1,2,3}, x'_{1,2,3} <0$ and $x_{4,5,6},x'_{4,5,6}>0$. The integral \eqn{zab} gives
\be {\cal Z} = \left[\left(\frac{s_{63}s_{6'3'}}{s_{63'}s_{6'3}}\right)^4\prod_{a=4}^5\left(\frac{s_{a3'}s_{a'3}}{s_{a3}s_{a'3'}}\right)^2 \prod_{b=1}^2\left(\frac{s_{b6'}s_{b'6}}{s_{b6}s_{b'6'}}\right)^2   \prod_{c=4}^5\prod_{d=1}^2\left(\frac{s_{cd}s_{c'd'}}{s_{cd'}s_{c'd}}   \right)\right]^{2/5}\ .\ee
We should now multiply this by the free correlators, given by 
\be
{\rm propagators} = \frac{1}{(2\pi)^6}\frac{s_{12}s_{1'2'}s_{36'}s_{63'}s_{45}s_{4'5'}}{s_{36}s_{3'6'}s_{33'}s_{66'}}\prod_{a=1}^2\prod_{b=4}^5\frac{s_{ab'}s_{a'b}}{s_{ab}s_{a'b'}}\prod_{c=1}^2\prod_{d=1}^2\frac{1}{s_{cd'}}\prod_{e=4}^5\prod_{f=4}^5\frac{1}{s_{ef'}}\ .\nn
\ee
Taking the clustering limit $t_i\gg t'_j$ allows us to read off the correlation function from the product as
%
\be
\langle \Theta_{11\bar{2}}(t_i,x_i)\rangle=\frac{1}{(2\pi)^3}\frac{s_{12}s_{45}}{s_{36}}\prod_{e=1}^2\prod_{f=4}^5\frac{1}{s_{ef}^{3/5}}\left[s_{63}^4\prod_{a=4}^5\frac{1}{s_{a3}^2} \prod_{b=1}^2\frac{1}{s_{b6}^2} \right]^{2/5}
\ee
where, as in previous examples, there is the possibility of an overall phase. 

\para
We can do the same for the second scattering process in \eqn{twoprocess}. This time we consider the operator
\be \Theta_{122}(t_i,x_i) = \psi_1(t_1,x_1)\psi_2(t_2,x_2)\psi_2(t_3,x_3)\psi_1(t_4,x_4)\psi_2(t_5,x_5)\psi_2(t_6,x_6)\ .\ee
Again, we start by computing the 12-point function $\langle \Theta_{122}(t_i,x_i)\Theta_{122}^\dagger(t'_i,x'_i)\rangle$ where we take $x_{1,2,3}, x'_{1,2,3} <0$ and $x_{4,5,6},x'_{4,5,6}>0$. Now, the integral \eqn{zab} gives
\be {\cal Z} = \left[\left(\frac{s_{14'}s_{1'4}}{s_{14}s_{1'4'}}\right)\prod_{a=5}^6\left(\frac{s_{1'a}s_{1a'}}{s_{1a}s_{1'a'}}\right)^2 \prod_{b=2}^3\left(\frac{s_{4'b}s_{4b'}}{s_{4b}s_{4'b'}}\right)^2   \prod_{c=2}^3\prod_{d=5}^6\left(\frac{s_{c'd}s_{cd'}}{s_{cd}s_{c'd'}}   \right)^4\right]^{2/5}\ .\ee
The  free propagators are 
\be
{\rm propagators} = \frac{1}{(2\pi)^6}\frac{s_{14}s_{1'4'}s_{56}s_{5'6'}s_{23}s_{2'3'}}{s_{14'}s_{1'4}s_{56'}s_{5'6}s_{23'}s_{2'3}}\prod_{a=2}^3\prod_{b=5}^6\frac{s_{ab}s_{a'b'}}{s_{ab'}s_{a'b}}\prod_{c=1}^6\frac{1}{s_{cc'}}\ .\nn
\ee
Taking the clustering limit $t_i\gg t'_j$ allows us to read off the correlation function from the product as
%
\be
\langle \Theta_{122}(t_i,x_i)\rangle=\frac{1}{(2\pi)^3}s_{14}s_{56}s_{23}\prod_{a=2}^3\prod_{b=5}^6 \frac{1}{s_{ab}^{3/5}} \left[\frac{1}{s_{14}}\prod_{a=5}^6\frac{1}{s_{1a}^2} \prod_{b=2}^3\frac{1}{s_{4b}^2} \right]^{2/5}\ .
\ee
Again, there is the possibility of an overall phase. 
 

\section{A Mod 2 Anomaly}\label{mzsec}


The simplest version of the fermion-rotor system \eqn{polact}, consisiting of $N$ fermions each with charge $q_i=1$, captures the s-wave scattering of $N$ Weyl fermions in $d=3+1$ dimensions, interacting with an $SU(2)$ 't Hooft-Polyakov monopole. That begs an interesting question. The 4d theory is only consistent when $N$ is even; otherwise it suffers a Witten anomaly \cite{wsu2}. It's natural to  wonder if perhaps there is some remnant of the Witten anomaly in the lower dimensional theory.
In this section, we confirm that this is indeed the case.

\para
We will see this in two different ways, the first by considering the theory on a spatial line (as we have so far), and the second by considering the theory on a spatial circle. In the first approach, we show that cluster decomposition arguments imply that fermionic (i.e. Grassmann-odd) operators have expectation values. This is reminiscent of the Witten anomaly, where instantons in theories that suffer a Witten anomaly have an odd number of fermionic zero modes.

\para
In the second method, on a spatial circle, we will show that the fermion-rotor system has an odd number of Majorana zero modes when there is a Witten anomaly. We discuss this in Section \ref{circlesec}, where we also give the more general statement for the model with an arbitrary number of rotors and arbitrary charges.

\para
To illustrate why the theory with $N$ odd is worrisome, we consider the simple case of $N=1$. We compute the correlator $\langle \psi(t,x) e^{i\alpha(\tau)/2} \,\psi^\dagger(t',x') e^{i\alpha(\tau')/2}\rangle$ for $x>0$ and $x'<0$ (Note that both rotor insertions have $e^{+i\alpha/2}$, rather than $e^{\pm i\alpha/2}$). It is straightforward to compute this correlator using the methods of Section \ref{corrsec}. It is independent of the IR cut-off $\mu$ but, in contrast to the correlation functions that we computed previously, it is UV divergent. This reflects the fact that $e^{i\alpha(\tau)/2}$ is not playing the role of a twist operator here: instead it excites the rotor.

\para
The UV divergence is an artefact of working with the integral \eqn{zab} that is derived in the  low-energy limit $E\ll 1/I$, with $I$ the moment of inertia of the rotor. If we include the original kinetic term for the rotor, as in \eqn{zwithi}, then the integral is fully convergent. If we just care about the scaling, then we can mimic this by imposing a UV cut-off on the integral \eqn{zab} given by  $\Lambda = 1/I$.

\para
The correlator factorises under cluster decomposition. We scale the insertion times as  $|t-t'|\sim T$ and $|\tau-\tau'|\sim T$ and take the limit $T\ra \infty$ to find, up to numerical factors, 
\be \langle \psi(t,x) e^{i\alpha(\tau)/2} \,\psi^\dagger(t',x') e^{i\alpha(\tau')/2}\rangle \,\sim \,\frac{I}{(s-\tau)(s'-\tau')}\ .\ee
Cluster decomposition then tells us that
\be \langle \psi(t,x) e^{i\alpha(\tau)/2}\rangle\, \sim\, \frac{\sqrt{I}}{s-\tau}\ .\ee
This is the promised expectation value for a Grassmann-odd field. It is the first sign that something is fishy. Similar behaviour occurs for all theories with $N$ odd, but not for those with $N$ even.

\subsection{The Fermion-Rotor System on a Circle}\label{circlesec}

We can get a more precise handle on the anomaly of this theory if we compactify on a spatial circle, with periodic boundary conditions on the fermion. Now the  global $U(1)$ symmetry \eqn{symf} associated to the rotor disappears, with the asymtotic form of $f(x)$ incompatible with periodic boundary conditions. 

\para
It's  unusual to lose a global symmetry when compactifying, but it's familiar for the global, or asymptotic, part of a gauge symmetry  which also disappears when you put the theory on compact spatial manifold. Indeed, in the reduction from the monopole, the $U(1)$ symmetry \eqn{symf} does descend from the global part of the $U(1)$ gauge symmetry in $d=3+1$.

\para
Now we can ask: what is the effect of the rotor on the low-energy spectrum of the theory? The answer follows immediately from our analysis in Section \ref{frsec} where we showed that the rotor acts as a twist operator. The low-energy theory on a circle is simply free fermions, but with appropriately twisted boundary conditions.
In terms of bosonised fields $\chi_i$, these twisted boundary conditions on a circle of radius $R$ are
\be \chi_i(x+2\pi R) = {\cal R}_{ij} \chi_i(x) \ \ee
where the rotation matrix ${\cal R}_{ij}$ was defined in \eqn{rij}. For the theory with $N$ fermions of charge $q_i=1$, the rotation matrix is
\be {\cal R}_{ij} = \delta_{ij} - \frac{2}{N}\ .\label{dyonrij}\ee
We can now see what is wrong with the theory when $N$ is odd. This is simplest when we have just a single $N=1$ fermion. In this case, the twisted boundary condition becomes
\be \psi(x+2\pi R) = \psi^\dagger(x)\ .\ee
If we decompose the fermion $\psi$ in terms of Majoranas, and write $\psi=\lambda_1 + i\lambda_2$, then $\lambda_1$ has periodic boundary conditions while $\lambda_2$ has anti-periodic boundary conditions. That means that $\lambda_1$ has a zero mode on the circle, while $\lambda_2$ does not. But a theory with a single (or, indeed, odd) number of Majorana zero modes is a notoriously subtle object. A naive computation of the Euclidean partition function gives $\sqrt{2}$, which is clearly does not carry the usual interpretation of the dimension of the Hilbert space. See \cite{dgg2,w,seiberg,freed} for recent discussions of this simple, but confusing, system. 

\para
The fermion-rotor system with any odd $N$ is plagued with the same issue. This follows from the work of \cite{philip1,philip2}, where it was shown that the boundary states specified by the rotation matrix ${\cal R}_{ij}$ fall into one of two classes that were termed  {\it vector} and {\it axial}. For a system on an interval, the boundary conditions on each end must lie in the same class to avoid an odd number of fermion zero modes. For the unfolded, chiral theory on a circle, only the vector-like class will avoid an odd number of zero modes. It was shown in \cite{philip1} that the boundary condition \eqn{dyonrij} lies in the vector class when $N$ is even and the axial class when $N$ is odd.

\para
The upshot is that the Witten anomaly in $d=3+1$ dimensions descends to the mod 2 anomaly $d=1+1$ dimensions which simply counts the number of Majorana zero modes. 

\para
This same connection between the Witten anomaly and the mod 2 anomaly of Majorana modes has been previously noted in a closely related context \cite{john,yuji}. These papers considered fermion bound states, rather than the chiral scattering states under discussion here. There it was also noted that monopoles in theories that suffer a Witten anomaly have an odd number of Majorana zero modes. 

\para
The same story applies to the more general class of models described in Section \ref{chargesec}. The symmetries of these models are captured in the rotation matrix ${\cal R}_{ij}$ defined in \eqn{rij}, and a recipe was provided in \cite{philip1} to determine the class (vector or axial) of any such ${\cal R}_{ij}$. The theory on a spatial circle has an even number of Majorana zero modes only when the ${\cal R}_{ij}$ lies in the vector class.

\para
As an example, we can consider the 3450 model, which  means that there is a $U(1)$ global symmetry under which the incoming fermions have charges 5 and 0, and the outgoing fermions have charges 3 and 4. There are different ways to construct this. One way, already mentioned in Section \ref{chargesec}, is to consider two fermions coupled to a single rotor with charges  $q_i=(1,2)$. In the notation of \eqn{ccbar}, the  3450 symmetry arises from the linear combination $- Q_{1i} +2Q_{2i} = (5,0)$ and $- \bar{Q}_{1i} + 2\bar{Q}_{2i} = (3,-4)$, with the corresponding rotation matrix

\be {\cal R} = \frac{1}{5}\left(\begin{array}{cc}  3 \ & -4\\ -4\ & -3\end{array}\right)\ .\ee
From the diagnostic presented in \cite{philip1}, this lies in the axial class: this model has an odd number of Majorana zero modes. One can show using cluster decomposition that there is a  non-vanishing Grassmann-odd expectation value $\langle \psi_1\psi_2\psi_2e^{5i\alpha/2}\rangle\neq 0$. 

\para
There is a straightforward way to construct a 3450 model with an even number of Majorana modes. We again take two fermions coupled to a single rotor, this time with $q_i=(1,3)$. The 3450 symmetry, no longer faithfully acting,  now arises from the linear combination $-Q_{1i} + 3 Q_{2i} = (10,9)$ and $-\bar{Q}_{1i} + 3\bar{Q}_{2i} = (8,-6)$.  This time the rotation matrix is
\be {\cal R} = \frac{1}{5}\left(\begin{array}{cc}  4 \ & -3\\ -3\ & -4\end{array}\right)\ .\ee
From \cite{philip1}, this lies in the vector class, and the theory on the circle has an even number of Majorana zero modes. Correspondingly, the operator that gets an expectation is the Grassmann-even $\langle\psi_1\psi_3\psi_3\psi_3 e^{5i\alpha}\rangle\neq 0$.

\section*{Acknowledgements} We're grateful to Philip Boyle Smith for useful discussions. This work was supported by the STFC grant ST/X000664/1, STFC grant ST/X000761/1, and a Simons Investigator Award. For the purpose of
open access, the author has applied a Creative Commons Attribution (CC BY) licence to any Author Accepted Manuscript version arising from this submission.

\appendix
\section{Appendix: The Rotor Partition Function}\label{appa}

 Our goal in this appendix is to compute the Lorentzian partition function for the action \eqn{oneqact} and reproduce the correlation functions given in  \eqn{fcorr} and \eqn{zab}. We start by introducing fermionic sources $\xi$ and $\xi^\dagger$, so the action reads
 \be S = \int d^2x \ i\sum_{i=1}^N\Big(\psi_i^\dagger {\cal D}_i \psi_i +\xi^\dagger_i\psi_i+{\psi}^\dagger_i\xi_i\Big)+ \int dt\ \frac{I}{2} \dot{\alpha}^2\ .\ee
with 
\be {\cal D}_i =\partial_++ iq_{i} \alpha(t) f'(x)\ .\ee
The partition function is then
\be Z[\xi,\xi^\dagger] &=& \int {\cal D}\psi{\cal D}\psi^\dagger  {\cal D}\alpha \ e^{iS}\nn\\ &=&\int {\cal D}\alpha\ \left(\prod_{i=1}^N {\rm det}\,{\cal D}_i\right)\,  \exp\left(i\int dt\ \frac{I}{2}\dot{\alpha}^2\right)\, \exp\left(\sum_{i=1}^N \int d^2x\ \xi_i^\dagger {\cal D}_i^{-1}\xi_i\right)\ .\ee
We need to do three things: compute the propagator ${\cal D}_i^{-1}$; compute the determinant ${\rm det}\,{\cal D}_i$; and integrate out the rotor degree of freedom $\alpha(t)$. We now do each of these in turn.

\para
First the propagator. In the absence of the rotor, we have a free chiral fermion with usual propagator
\be G_0(t-t',x-x')= \bra{t,x}\partial_+^{-1}\ket{t',x'} = \frac{1}{2\pi i} \frac{t-t' + x-x'}{(t-t')^2-(x-x')^2 - i\epsilon}\ .\ee
But it's a simple matter to absorb the effect of the  rotor into a phase shift of the propagator. We have
\be {\cal D}_i^{-1} = e^{-iq_i \lambda} \partial_+^{-1} e^{iq_i\lambda} \ee
where
\be   \lambda(t,x) = \partial_+^{-1} \alpha(t) f'(x)  = \int dt' dx'\ G_0(t-t',x-x') \, \alpha(t') f'(x')  \ .\label{lamtx}\ee
 In Fourier space, we have
 \be \lambda(t,x)&=&\int dx'\frac{d\omega}{2\pi}\frac{dk}{2\pi} \ e^{-i\omega t}\alpha(\omega) f'(x')\, \frac{-i(k+\omega)}{k^2 - \omega^2 - i\epsilon} e^{ik(x-x')}
 \nn\\&=& \int dx'\frac{d\omega}{2\pi}\ e^{-i\omega t}\alpha(\omega) f'(x')\big[\theta(x-x')\theta(\omega) -\theta(x'-x)\theta(-\omega)\big] e^{i\omega(x-x')}\nn\\
 &=& \int dx'\frac{d\omega}{2\pi}\ \theta\big(\omega(x-x')\big) {\rm sign}(\omega)\,e^{-i\omega (t-(x-x'))}\alpha(\omega) f'(x')
 \ .\ee
At this point, we take the  small core limit,  $f'(x) \ra \delta(x)$. We have the phase shift
\be \lambda(t,x) &=& \int \frac{d\omega}{2\pi}\  \theta(\omega x) {\rm sign}(\omega)\,e^{-i\omega(t-x)} \alpha(\omega) \nn\\
&=& \int_0^\infty \frac{d\omega}{2\pi}\  \Big[ \theta(x) \alpha(\omega) e^{-i\omega(t-x)} - \theta(-x) \alpha(-\omega) e^{i\omega(t-x)}\Big]\ .\label{lambda8}\ee
This is the first of our three results. The partition function becomes
\be {\cal Z}[\xi,\xi^\dagger] &=& \int {\cal D}\alpha \, \left(\prod_{i=1}^N{\rm det}\,{\cal D}_i\right) \, \exp\left(i\int dt\ \frac{I}{2}\dot{\alpha}^2\right)\\ &&\ \ \  \times\, \exp\left(\sum_{i=1}^N \int d^2x\,d^2x'\ \xi_i^\dagger(t,x) e^{-iq_i\lambda(t,x)} G_0(t-t',x-x')e^{iq_i\lambda(t',x')}\xi_i(t',x')\right) \ .\nn\ee
Differentiating with respect to the sources $\xi$ and $\xi^\dagger$ then gives us the intermediate expression for fermion correlators
\be \Big< \prod_{j=1}^n\psi_{i_j}(t_j,x_j)\prod_{k=1}^{n'} \psi^\dagger_{i'_k}(t'_k,x'_k) \Big> = (\mbox{free correlators})\times {\cal Z}\ee
with 
\be {\cal Z} =  \int {\cal D}\alpha \ \left(\prod_{i=1}^N{\rm det}\,{\cal D}_i\right)   \exp\left(i\int dt\ \frac{I}{2}\dot{\alpha}^2\right) \exp\left(-i\int_0^\infty \frac{d\omega}{2\pi}\ \big( A\,\alpha(\omega) -B\,\alpha(-\omega)\big)\right)\ .\nn\ee
Here the two functions $A$ and $B$ can be extracted from the expression \eqn{lambda8} for $\lambda$ and are given by
\be A &=& \sum_{j=1}^n q_{i_j}\theta(x_j) e^{-i\omega (t_j-x_j)} -  \sum_{k=1}^{n'} q_{i'_k}\theta(x'_k) e^{-i\omega (t'_k-x'_k)}\nn\\
 B &=& \sum_{j=1}^n q_{i_j}\theta(-x_j) e^{+i\omega (t_j-x_j)} -  \sum_{k=1}^{n'} q_{i'_k}\theta(-x'_k) e^{+i\omega (t'_k-x'_k)}
 \ .\label{a11}\ee
This coincides with the expression \eqn{aandb} given in the main text. 

\para
Next, we turn to the determinant. For the operator ${\cal D} = \partial_+ + iq\alpha(t)f'(x)$, we have
\be {\rm det}\,{\cal D} = \exp\log {\rm det}\,{\cal D} = \exp{\rm Tr}\log\left[\partial_+(1+i\lambda(t,x))\right]\ee
with $\lambda(t,x)$ defined in \eqn{lamtx}. Expanding the log, we can drop the constant term, while the term linear in $\lambda$ vanishes on account of CPT. The  non-trivial term arises at order $\lambda^2$ and, although it is not obvious from the expansion of the log, this is the exact result. This can be seen in a diagrammatic approach. (Alternatively, see the footnote below Eq.~(41) of \cite{joe}.) We have
\be {\rm det}\,{\cal D} = \exp\left(\frac{q^2}{2}\int d^2x\,d^2x'\ \alpha(t)\alpha(t')f'(x)f'(x')\,G_0^2(t-t',x-x')\right)\ .\ee
We evaluate this in Fourier space. After taking the small core limit, $f'(x) \ra \delta(x)$, we have
\be {\rm det}\,{\cal D} = \exp\left(-\frac{q^2}{4\pi}\int_0^\infty\frac{d\omega}{2\pi} \ \omega\alpha(\omega)\,\alpha(-\omega)\right)\ .\ee
Crucially, this term is linear in $\omega$, contrasting with the original rotor kinetic term which is of the form $I\omega^2$. Inlcuding them both, we have 
\be {\cal Z} &=& \int {\cal D}\alpha\ \exp\left(-\int_0^\infty\frac{d\omega}{2\pi} \left( \frac{\sum_i q_i^2}{4\pi}  \omega-iI\omega^2\right)\alpha(\omega)\,\alpha(-\omega)\right)\nn\\ &&\ \ \ \ \ \  \ \ \ \times\, \exp\left(-i\int_0^\infty \frac{d\omega}{2\pi}\ \big( A\,\alpha(\omega) -B\,\alpha(-\omega)\big)\right) \nn\ee
We're left with a  partition function that is quadratic in the rotor degree of freedom $\alpha(\omega)$, allowing us to easily perform the functional integral.
\be {\cal Z} &=& \exp\left(2 \int_0^\infty d\omega\ \frac{AB}{(\sum_iq_i^2)\omega - 4\pi iI\omega^2}\right) .\label{zwithi}\ee
The $I\omega^2$ term ensures that the integral is convergent, even if (as is sometimes the case) $AB$ has a constant term. For most applications in this paper, there is no constant term in $AB$ and in this case we may take the  low-energy limit, $\omega \ll 1/I$, allowing us to  drop the original kinetic term. We're left with
\be {\cal Z} &=& \exp\left(\frac{2}{\sum_iq_i^2} \int_0^\infty \frac{d\omega}{\omega} \ AB\right) .\ee
This is the result advertised in \eqn{fcorr} and \eqn{zab}.

\end{document}